\documentclass[journal]{IEEEtran}

\usepackage{cite}
\usepackage{amsmath}
\usepackage{amssymb}
\usepackage{amsfonts}
\usepackage{graphicx}
\usepackage{textcomp}
\usepackage{xcolor}
\usepackage{array}
\usepackage{subfigure}
\usepackage[font=normalsize]{caption}
\usepackage{stfloats}
\usepackage{url}
\usepackage{verbatim}
\usepackage[export]{adjustbox}
\usepackage{ntheorem}
\usepackage{longtable}
\usepackage{booktabs}
\usepackage{makecell}
\usepackage{mwe}
\usepackage{graphbox}
\usepackage{enumitem}
\usepackage{ragged2e}
\usepackage{multicol} 
\usepackage[colorlinks, linkcolor=black, anchorcolor=black, citecolor=blue]{hyperref}
\usepackage{comment}
\usepackage{color}
\usepackage{bbding} 
\usepackage{diagbox} 
\usepackage{float}
\usepackage{multirow}
\theorembodyfont{\upshape}
\newtheorem{theorem}{Theorem}[section] 

\newtheorem{definition}[theorem]{Definition}

\makeatletter
\renewtheoremstyle{plain}
{\item{\theorem@headerfont ##1\ ##2\theorem@separator}~}
{\item{\theorem@headerfont ##1\ ##2\ (##3)\theorem@separator}~}
\makeatother
\def\BibTeX{{\rm B\kern-.05em{\sc i\kern-.025em b}\kern-.08em
		T\kern-.1667em\lower.7ex\hbox{E}\kern-.125emX}}
\usepackage{balance}
\graphicspath{{QUDM/EPS/}}
\usepackage{lipsum}


\usepackage{setspace}
\graphicspath{{}}
\setenumerate[1]{itemsep=0pt,partopsep=0pt,parsep=\parskip,topsep=0pt}
\setitemize[1]{itemsep=0pt,partopsep=0pt,parsep=\parskip,topsep=0pt}
\setdescription{itemsep=0pt,partopsep=0pt,parsep=\parskip,topsep=0pt}

\usepackage[font=normalsize]{caption}
\usepackage[ruled,vlined]{algorithm2e}
\begin{document}
\title{PGKET: A Photonic Gaussian Kernel Enhanced Transformer}	
\author{Ren-Xin Zhao
		
\IEEEcompsocitemizethanks{
\IEEEcompsocthanksitem{
}
\IEEEcompsocthanksitem{Ren-Xin Zhao is with the School of Computer Science and Engineering, Central South Univerisity, China, Changsha, 410083 and visiting the School of Computer Science and Statistics, Trinity College Dublin, Ireland, Dublin, D02 PN40 (Email: renxin\_zhao@alu.hdu.edu.cn). 
}
}
}

\maketitle
	
\begin{abstract}
Self-Attention Mechanisms (SAMs) enhance model performance by extracting key information but are inefficient when dealing with long sequences. To this end, a photonic Gaussian Kernel Enhanced Transformer (PGKET) is proposed, based on the Photonic Gaussian Kernel Self-Attention Mechanism (PGKSAM). The PGKSAM calculates the Photonic Gaussian Kernel Self-Attention Score (PGKSAS) using photon interferometry and superposition to process multiple inputs in parallel. Experimental results show that PGKET outperforms some state-of-the-art transformers in multi-classification tasks on MedMNIST v2 and CIFAR-10, and is expected to improve performance in complex tasks and accelerate the convergence of Photonic Computing (PC) and machine learning.
\end{abstract}


\section{Introduction}
\IEEEPARstart{I}{n} recent years, Transformers have progressed remarkably relying on SAMs \cite{0.0,0.1,0.2,0.3}. The SAM, first proposed in 2017, generates new sequences by calculating self-attention scores of sequences, captures critical information effectively, and reduces the dependence on external data, thus significantly improving the efficiency of models \cite{0.4}. With these merits, SAMs have become increasingly popular in areas such as computer vision \cite{0.5,0.6,0.7}. One prominent instance is that a SAM successfully reduces the floating point of operations by 52.0\% and improves Top-1 accuracy by 0.6\% on DeiT-B \cite{0.8}. Despite the power of SAMs, the inefficiency of many SAMs in handling long sequences has become a recognized challenge \cite{0.9}. Fortunately, kernel methods can effectively overcome this dilemma. In 2019, a SAM based on kernel methods was first introduced, paving the way for designing attention using a larger representation space \cite{0.10}. In 2021, Gaussian kernel methods replaced softmax in SAMs, significantly reducing computational cost while improving accuracy on long sequence data without windowing techniques \cite{0.11,0.12,0.13}. These cases sufficiently prove that SAMs incorporating kernel methods have broken the shackles of data representation and computational complexity, thus opening up new paths for their development. Given the successful use of kernel methods in SAMs, is it possible to further enhance their performance? PC is expected to provide the answer.

PC is an emerging technology that manipulates photonic states in photonic systems for information processing, which is rapidly spreading in the fields of communication \cite{0.13.1} and machine learning \cite{0.13.2}. It embodies multiple advantages. Firstly, photons are able to transfer information in multiple paths simultaneously to achieve efficient parallel computation, which greatly improves computational efficiency and is particularly suitable for large-scale data processing \cite{0.14,0.15,0.16}. Second, manipulating photonic devices can produce complex and sophisticated photonic gate operations, making the transition between photonic states more flexible \cite{0.17,0.18}. Furthermore, the low-loss characteristics of photons as information carriers enable PC systems to effectively reduce energy consumption when performing high-speed calculations \cite{0.19,0.20}. Finally, the interference and entanglement properties equip photons with the ability to work together in a non-classical way, thus showing great potential to outperform conventional computers in certain computational problems \cite{0.21}. In a word, the successful application of kernel methods and the multiple advantages of PC provide theoretical and practical guidance for the innovations in this paper. Specifically, the contributions of this paper are as follows:

\begin{itemize}
 \item A PGKSAM-based PGKET is proposed to combine the strengths of both PC and kernel methods to break through the computational limitations of classical SAM.
 \item Experiments show that the PGKET outperforms conventional Transformers by virtue of PC in binary classification tasks on MNIST, Fashion MNIST and CIFAR-10.
\end{itemize}

 The rest of the paper is structured as follows: in Section \ref{sec2}, an overview of the Gaussian Kernel Self-Attention Mechanism (GKSAM) and PC is given to provide a theoretical basis for the creation of PGKSAM. In Section \ref{sec3}, the PGKSAM and the PGKET framework are described in detail. In Section \ref{sec4}, three experiments are conducted and a series of meaningful findings are obtained. In the end, conclusions are drawn based on the findings of the previous sections.

\section{Preliminaries}\label{sec2}
In this section, a brief overview of the GKSAM and PC is given.

\subsection{Gaussian Kernel Self-Attention Mechanism}
A GKSAM is proposed in 2021 to substitute softmax in SAM with a full covariance trainable matrix \cite{0.13}. Specifically, for an input sequence
\begin{equation}\label{input}
 \setlength{\abovedisplayskip}{3pt}\setlength{\belowdisplayskip}{3pt}
 \mathbf{In}=\{{{X}_{i}}\}_{i=1}^{n}
\end{equation} 
with ${{X}_{i}}=[x_1^i, \cdots, x_d^i]^{\text{T}}\in {{\mathbb{R}}^{1\times d}}$, the GKSAM is set as
\begin{equation}\label{GKSAM}
\setlength{\abovedisplayskip}{3pt}\setlength{\belowdisplayskip}{3pt}
\text{GKSAM}_i:=\sum\limits_{j}{A(i,j){{V}_{j}}}.
\end{equation} 
The most important part in \eqref{GKSAM} is the self-attention score
\begin{equation}\label{score}
\setlength{\abovedisplayskip}{3pt}\setlength{\belowdisplayskip}{3pt}
A(i,j):=\frac{1}{Z}\exp \left[ \frac{-1}{2}{{({{X}_{i}}-{{X}_{j}})}^{\text{T}}}\Gamma ({{X}_{i}}-{{X}_{j}}) \right],
\end{equation}
where the normalization factor
\begin{equation}\label{z}
\setlength{\abovedisplayskip}{3pt}\setlength{\belowdisplayskip}{3pt}
Z=\sum\limits_{j}{\exp \left[ \frac{-1}{2}{{({{X}_{i}}-{{X}_{j}})}^{\text{T}}}\Gamma ({{X}_{i}}-{{X}_{j}}) \right]}
\end{equation} ensures that the sum of self-attention scores equals 1. The full covariance trainable matrix $\Gamma$ in \eqref{score} and \eqref{z} measures the statistical relationship between the query transformation matrix $W^Q$ and the key transformation matrix $W^K$.
In the end, the value vector
\begin{equation}\label{Vj}
\setlength{\abovedisplayskip}{3pt}\setlength{\belowdisplayskip}{3pt}
{{V}_{j}}:={{X}_{j}}{{W}^{V}},
\end{equation} 
where $W^V$ is the trainable value transformation matrix. The authors point out that the GKSAM relies on the difference ${{X}_{i}}-{{X}_{j}}$ between the input features and therefore has translation invariance, while removing the energy term in SAM to simplify the computation.

\subsection{Photonic Computing}
PC takes photons as information carriers for processing and analysis. In PC, the smallest unit of information is a qumode. Its photonic state 
\begin{equation}\label{qumode}
\setlength{\abovedisplayskip}{3pt}\setlength{\belowdisplayskip}{3pt}
|\psi \rangle =\sum\limits_{k=0}^{\infty }{{{c}_{k}}}|k\rangle \in {{\mathcal{H}}^{\infty }},
\end{equation} 
where $c_k$, the probability amplitude of the ground state $|k\rangle $, satisfies the normalisation condition $\sum\nolimits_{k}{|{{c}_{k}}{{|}^{2}}}=1$. ${{\mathcal{H}}^{\infty }}$ denotes the infinite dimensional Hilbert space. In practice, an infinite dimensional space is usually not considered, but a suitable cutoff dimension $N$ is chosen as the maximum photon number to allow \eqref{qumode} to be approximated as
\begin{equation}\label{qumode1}
\setlength{\abovedisplayskip}{3pt}\setlength{\belowdisplayskip}{3pt}
 |\hat{\psi }\rangle =\sum\limits_{k=0}^{N}{{{c}_{k}}}|k\rangle .
\end{equation} 
The evolution of \eqref{qumode1} relies on photonic gates. 
In this paper, beamsplitter gates
\begin{equation}\label{BS}
\setlength{\abovedisplayskip}{3pt}\setlength{\belowdisplayskip}{3pt}
 BS(\theta, \phi)=\exp(\theta, \phi)=\exp[\theta (e^{i \phi} a_1 a_2^\dagger -e^{-i \phi} a_1^\dagger a_2)]
\end{equation} 
and displacement gates 
\begin{equation}\label{D}
\setlength{\abovedisplayskip}{3pt}\setlength{\belowdisplayskip}{3pt}
 D(\alpha)=\exp(\alpha a^\dagger -\alpha^* a)
\end{equation} 
in Tab. \ref{Notations} are used, where $\hat{a}^{\dagger}$ and $\hat{a}$ are the creation and annihilation operators of the photon, respectively. 
In addition, there are three properties for \eqref{D} \cite{1.0,1.1}: 
(1) By the unitary nature of \eqref{D}, then
\begin{equation}\label{properties1}
			\setlength{\abovedisplayskip}{3pt}
	\setlength{\belowdisplayskip}{3pt}
	{{D}^{\dagger }}(\alpha)=D(-\alpha).
\end{equation}
(2) For two arbitrary inputs $a$ and $b$, one obtains
\begin{equation}\label{properties2}
			\setlength{\abovedisplayskip}{3pt}
	\setlength{\belowdisplayskip}{3pt}
	D(a)D(b)=D(a+b)\exp [(b{{a}^{*}}-{{b}^{*}}a)/2].
\end{equation} 
(3) When \eqref{D} acts on the ground state $|0\rangle$, 
\begin{equation}\label{properties4}
			\setlength{\abovedisplayskip}{3pt}
	\setlength{\belowdisplayskip}{3pt}
	D(z)|0\rangle =\exp \left( -\frac{|z{{|}^{2}}}{2} \right)\sum\limits_{b=0}^{\infty }{\frac{{{z}^{b}}}{\sqrt{b!}}|b\rangle } .
\end{equation} 
Finally, measurements of photonic states are usually carried out by photonic detectors, which are capable of counting the number of photons in each qumode to obtain information about the quantum state. Through the preparation, evolution and measurement of these qumodes, PC is able to implement complex quantum algorithms and information processing tasks, showing its potential for a wide range of applications.
\begin{table}[h!]
	\setlength{\abovedisplayskip}{3pt}
	\setlength{\belowdisplayskip}{3pt}
	\def\tablename{Tab.}
	\centering
	\caption{Photonic Gates}
	\label{Notations}
	\small
	\begin{tabular}{@{}lcc@{}}
		\toprule
		Name of Photonic Gate &\makecell[c]{Mathematical\\Symbol} & \makecell[c]{Symbol of\\Photonic Gates} \\ \midrule 
		
		Beamsplitter gate & $BS(\theta, \phi)$& 	\includegraphics[align=c,scale=0.17]{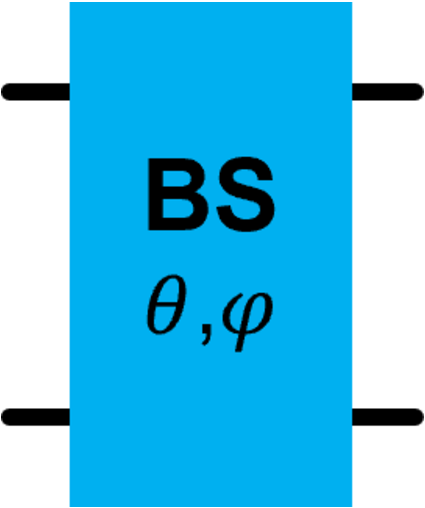} \\ 
		
		Displacement gate & $D(\alpha) $& 
		\includegraphics[align=c,scale=0.17]{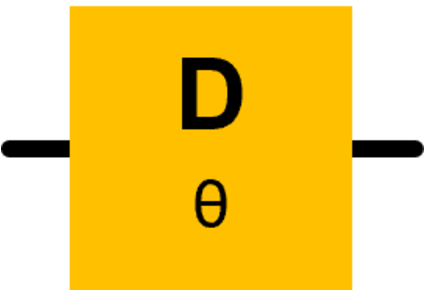} \\ 

		\bottomrule
	\end{tabular}%
\end{table}

\section{Photonic Gaussian Kernel Enhanced Transformer}\label{sec3}
In this section, we present the PGKSAM principle and the PGKET model.

\subsection{Photonic Gaussian Kernel Self-Attention Mechanism}

\begin{definition}\label{def1}
 The PGKAM 
\begin{equation}\label{PGKSAM}
\setlength{\abovedisplayskip}{3pt}\setlength{\belowdisplayskip}{3pt}
\text{PGKSA}{{\text{M}}_{i}}:=\sum\limits_{j}{\text{PGKSAS}(i,j){{V}_{j}}},
\end{equation} combines the efficiency of PC with the feature extraction advantages of the SAM, where the PGKSAS
\begin{equation}\label{PGKSAS}
\setlength{\abovedisplayskip}{3pt}\setlength{\belowdisplayskip}{3pt}
\text{PGKSAS}(i,j):=\exp (\mathbf{\Theta },\mathbf{f})\exp (-|{{X}_{i}}-{{X}_{j}}{{|}^{2}}/2) 
\end{equation} 
as the most important part in \eqref{PGKSAM} is estimated on the photonic processor.
$X_i$ and $X_j$ are from \eqref{input}. $\exp (\mathbf{\Theta },\mathbf{f})$ denotes a trainable shared weight matrix.
\end{definition}

\begin{figure}[h!]
 \centering
 \includegraphics[scale=0.27]{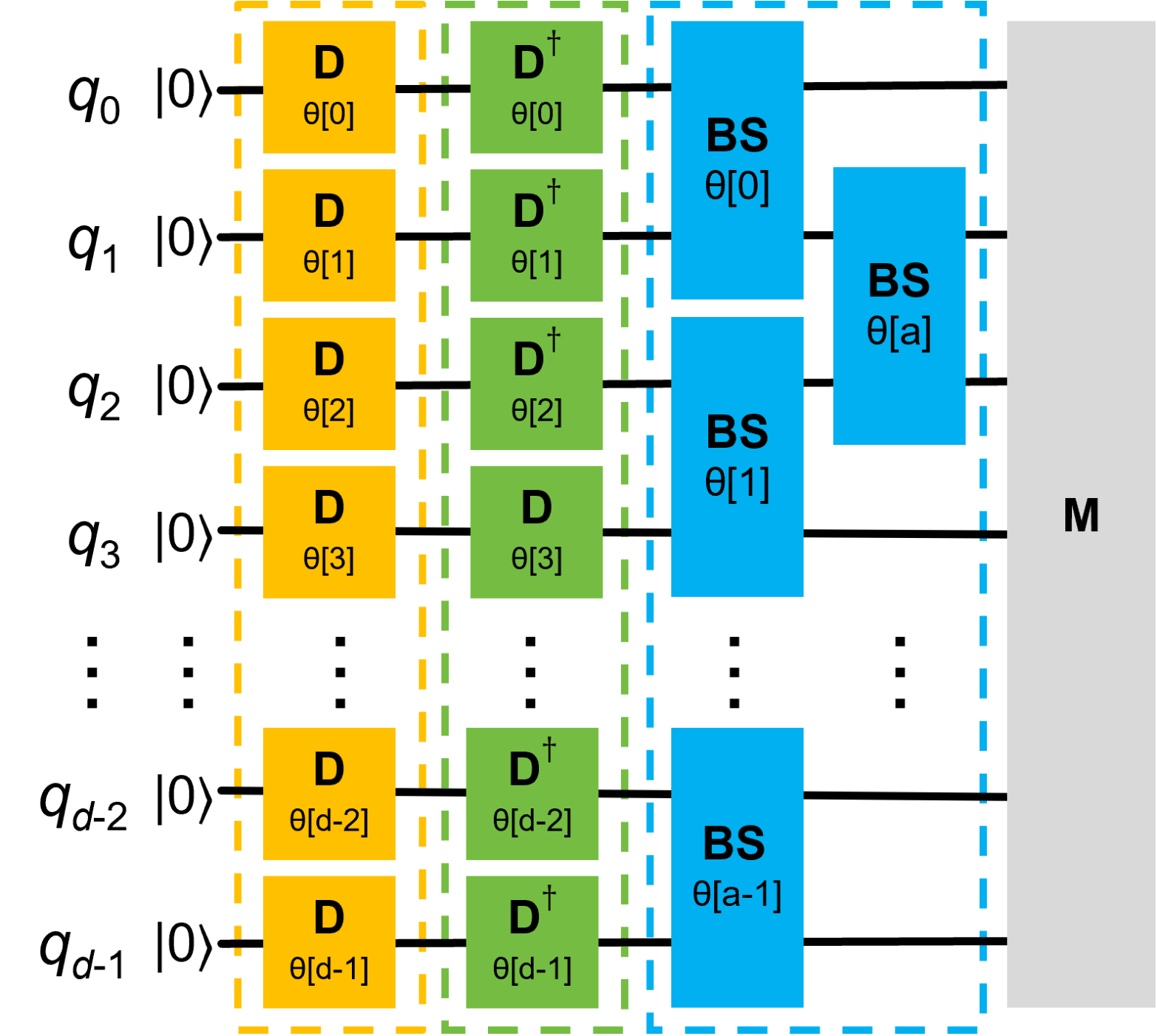}
 \caption{A PGKSAS photonic circuit}
 \label{circuit}
 
\vspace{10pt} 
 \centering
 \includegraphics[scale=0.3]{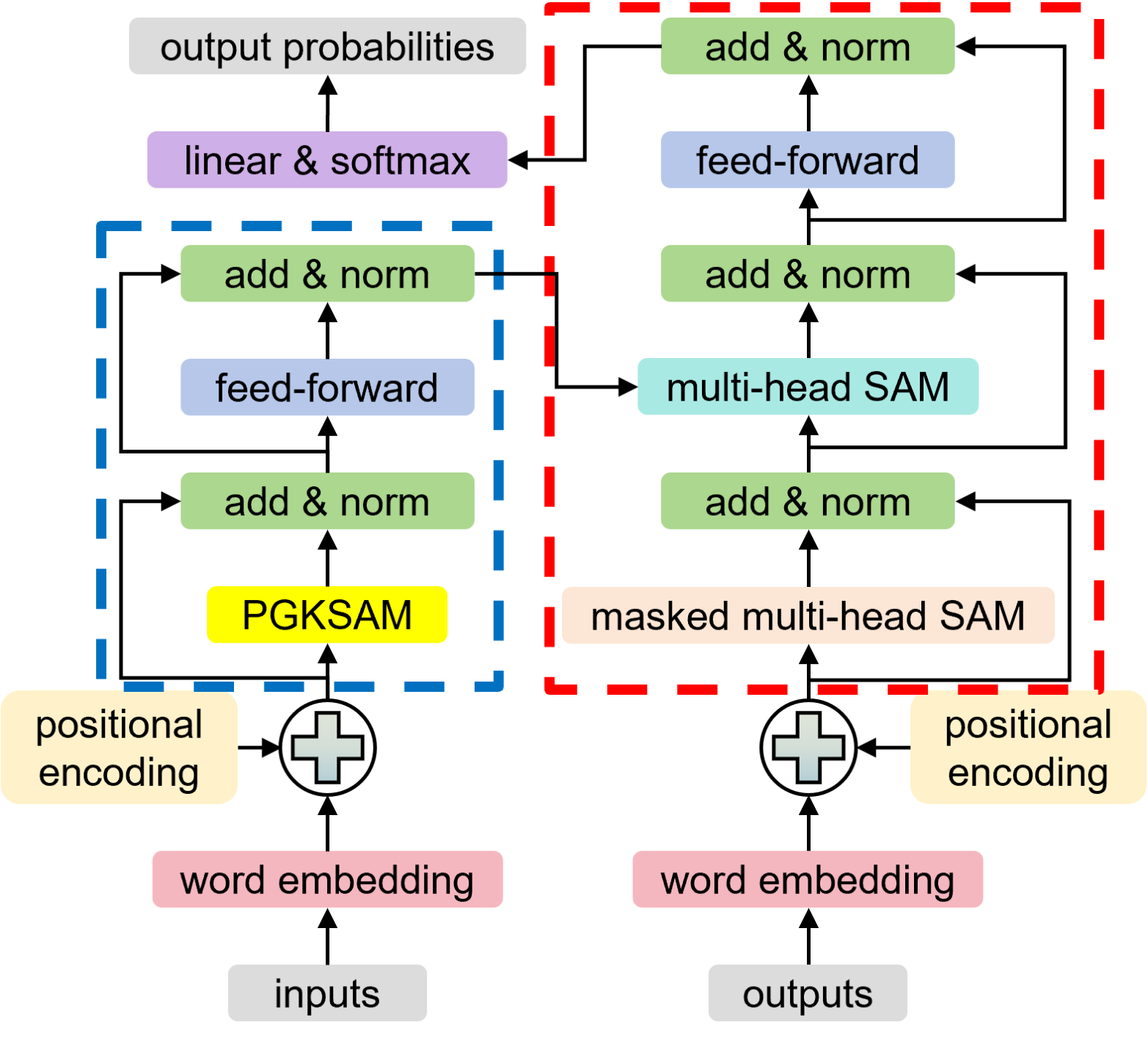}
 \caption{A PGKET framework: (1) Word embedding: Convert inputs and outputs into vector representations of fixed dimensions. (2) Positional encoding: Mark the relative position information between sequence elements. (3) PGKSAM: Optimize attention weights and similarity computation with a photonic circuit. (4) Add \& norm: Transmit the information deeper without loss, to enhance the fitting ability of the model, and prevent the learning process from abnormalities caused by too large or too small parameters. (5) Feed-forward: Nonlinearly capture complex patterns and relationships in the data. (6) Linear \& softmax: Map the high-dimensional representation of the model to the task-specific output space. (7) Masked muti-head SAM: Allow the decoder to make predictions based only on generated tokens. (8) Muti-head SAM: Let the token in the decoder side pay more attention to the corresponding token in the encoder. (9) Linear \& softmax: Transform to the matrix format required for the specific task.}
 \label{PGKET}
\end{figure}
With Definition \ref{def1}, the focus is on the implementation of \eqref{PGKSAS} on a photonic processor. Thus, we propose a photonic circuit for computing PGKSAS as shown in Fig. \ref{circuit}. 
Structurally, Fig. \ref{circuit} contains $d$ qumodes to correspond to the length of $X_i$ in \eqref{input}. 
$D(\alpha)$ in the orange dashed box and $D^\dagger(\alpha)$ in the green dashed box load $X_i$ and $X_j$ into the photonic system, i.e.,
\begin{equation}
\setlength{\abovedisplayskip}{3pt}\setlength{\belowdisplayskip}{3pt}
 \begin{aligned}[b]
 & {{X}_{i}}\to D({{X}_{i}})=\underset{t=1}{\overset{d}{\mathop{\otimes }}}\,D(x_{t}^{i}) \\ 
 & {{X}_{j}}\to {{D}^{\dagger }}({{X}_{j}})=\underset{t=1}{\overset{d}{\mathop{\otimes }}}\,{{D}^{\dagger }}(x_{t}^{j}).
\end{aligned}
\end{equation} 
In the blue dashed box, many $BS(\theta, \phi)$ are stacked into $\exp (\mathbf{\Theta },\mathbf{f})$ alternating according to Fig. \ref{circuit}. The gray box represents the measurement. The mathematical principle of Fig. \ref{circuit} is as follows: 

Step 1: the system is initially in the vacuum state
\begin{equation}
\setlength{\abovedisplayskip}{3pt}\setlength{\belowdisplayskip}{3pt}
 |\varphi\rangle = |\mathbf{0}\rangle =\underset{c=1}{\mathop{\overset{d}{\mathop{\otimes }}\,}}\,|0\rangle .
\end{equation}

Step 2: $D(\alpha)$ and $D^\dagger(\alpha)$ gates are applied:
\begin{equation}\label{step2}
\setlength{\abovedisplayskip}{3pt}\setlength{\belowdisplayskip}{3pt}
 |\varphi \rangle ={{D}^{\dagger }}({{X}_{j}})D({{X}_{i}})|\mathbf{0}\rangle .
\end{equation}
Combining \eqref{properties1} and \eqref{properties2}, \eqref{step2} is equal to
\begin{equation}\label{step3}
\setlength{\abovedisplayskip}{3pt}\setlength{\belowdisplayskip}{3pt}
 |\varphi \rangle =D({{X}_{i}}-{{X}_{j}})\exp (0)|\mathbf{0}\rangle .
\end{equation}
According to \eqref{properties4}, $\exp ( -|{{X}_{i}}-{{X}_{j}}{{|}^{2}}/2 ) $
in \eqref{PGKSAS} is further derived from \eqref{step3}.

Step 3: $\exp (\mathbf{\Theta },\mathbf{f})$ in \eqref{PGKSAS} is designed based on the arrangement rule of Fig. \ref{circuit}.

\subsection{Photonic Gaussian Kernel Enhanced Transformer}
A PGKET is proposed, whose core innovation lies in the optimization of the GKSAM computation using photonic circuits. Specifically, the PGKET fuses the flexibility of the GKSAM to capture similarities and non-linear relationships between data and PC to significantly accelerate computation and reduce energy consumption.

The concrete framework of PGKET is shown in Fig. \ref{PGKET}, and the functions of each sub-module have been introduced below. 
The workflow of Fig. \ref{PGKET} is dissected in detail next. First, the word embedding in the red box encodes the input and output as $\mathbf{raw_X}=\{{{x}_{i}}\}_{i=1}^{n}$ and $\mathbf{raw_Y}=\{{{y}_{i}}\}_{i=1}^{n}$, respectively. Then the positional encoding $p$ adds position information to $\mathbf{raw_X}$ and $\mathbf{raw_Y}$ to obtain 
\begin{equation}\label{pe1} 
\setlength{\abovedisplayskip}{3pt}\setlength{\belowdisplayskip}{3pt} 
\mathbf{In}=\mathbf{raw_X} + p=\{{{X}_{i}}\}_{i=1}^{n} \end{equation} and 
\begin{equation}\label{pe2} 
\setlength{\abovedisplayskip}{3pt}\setlength{\belowdisplayskip}{3pt} \mathbf{Out}=\mathbf{raw_Y} + p=\{{{Y}_{i}}\}_{i=1}^{n}. \end{equation}
Subsequently, \eqref{pe1} and \eqref{pe2} pass through the encoder represented by the blue dashed box and the decoder represented by the red dashed box respectively. In the blue dashed box, \eqref{PGKSAM} efficiently transforms \eqref{pe1} into 
\begin{equation}\label{out2} 
\setlength{\abovedisplayskip}{3pt}\setlength{\belowdisplayskip}{3pt} o_1=\{{{\text{PGKSAM}}_{i}}\}_{i=1}^{n} 
\end{equation}
on the photonic processor.
\eqref{pe1} and \eqref{out2} are added and layer normalized:
\begin{equation}\label{out3} 
\setlength{\abovedisplayskip}{3pt}\setlength{\belowdisplayskip}{3pt} {{o}_{2}}=\text{LayerNorm}(\textbf{In}+{{o}_{1}}).
\end{equation}
The feed-forward neural network processes \eqref{out3} to get \begin{equation}\label{out4} 
\setlength{\abovedisplayskip}{3pt}\setlength{\belowdisplayskip}{3pt} o_3=\text{ReLU}(W\cdot o_2+b),
\end{equation}
where \text{ReLU} is the activation function. $W$ is the weight matrix. $b$ is the bias term. Afterwards, \eqref{out3} and \eqref{out4} are again subjected to residual addition and layer normalization operations:
\begin{equation}\label{out5} 
\setlength{\abovedisplayskip}{3pt}\setlength{\belowdisplayskip}{3pt} {{o}_{4}}=\text{LayerNorm}({{o}_{2}}+{{o}_{3}}).
\end{equation}
In the red dashed box, most of the modules are the same as in the blue dashed box, leading to their working mechanism consistent with \eqref{out2}-\eqref{out5}, except that the input is \eqref{pe2}. Notably, the masked multi-head SAM 
\begin{equation}\label{out6} 
\setlength{\abovedisplayskip}{3pt}\setlength{\belowdisplayskip}{3pt} o_5= \text{softmax}[ \mathbf{Out}\cdot {{W}^{Q}}{{(\mathbf{Out}\cdot {{W}^{K}})}^{\text{T}}}/{\sqrt{d}} ]\mathbf{Out}\cdot {{W}^{V}}
\end{equation}
and the multi-head SAM 
\begin{equation}\label{out7} 
\setlength{\abovedisplayskip}{3pt}\setlength{\belowdisplayskip}{3pt} o_6= \text{softmax}[ \mathbf{Out}\cdot {{W}^{Q'}}{{(\mathbf{In}\cdot {{W}^{K'}})}^{\text{T}}}/{\sqrt{d}} ]\mathbf{In}\cdot {{W}^{V'}}
\end{equation}
are different from the PGKSAM, where $\text{softmax}$ is the activation function. $\sqrt{d}$ is a scaling factor. ${W}^{Q}$, ${W}^{Q'}$, ${W}^{K}$, ${W}^{K'}$, ${W}^{V}$ and ${W}^{V'}$ represent trainable transfer matrices. At this point, the PGKET has been fully introduced, laying the foundation for the experiment.

\section{Experiments}\label{sec4}
In this section, the performance of the PGKET is assessed on the Qutip and Pytorch platforms for MedMNIST v2 \cite{3.0} and CIFAR-10 \cite{3.1} multi-classification tasks. Specifically, the following three experiments are conducted.
\begin{itemize}
 \item \textbf{Experiment 1}: The performance of PGKET, HQViT \cite{3.0} and Quixer \cite{3.1} is evaluated and compared on MedMNIST v2 and CIFAR-10 datasets in 5-classification tasks in the absence of noise and with the same classical optimizer configuration.
\end{itemize}

\subsection{Datasets and Experimental Settings}
\subsubsection{Datasets}
MedMNIST v2 and CIFAR-10 are widely used datasets in biomedical and machine learning research. MedMNIST v2 has 12 standardized 2D biomedical imaging subsets such as PathMNIST, totaling about 708069 $28\times28$ pixel grayscale images. CIFAR-10 contains 60000 $32\times32$ RGB color images of 10 object categories, such as airplanes, divided into 50000 training and 10000 test samples. The selection of these two datasets is based on the following two reasons. (1) \textbf{Hardware constraints}: Although Xanadu has released a high-performance photonic computer named Aurora for paid use, international policies have hindered our access to these real photonic computers. As a result, PGKET cannot fully test its potential on real photonic hardware. This forces PGKET to run numerical simulations on classical servers with limited storage space for performance testing. (2) \textbf{Simulation limitations}: Processing more complex datasets such as ImageNet using PGKET severely exceeds the storage capacity of our classical servers (Google Colab), making large-scale simulations impossible. This is because the free version of Google Colab only provides 2 CPUs, 13GB of memory, and a GPU with an unspecified service time.

Based on the above two points, images from the top five categories in MedMNIST v2 (or CIFAR-10) are first extracted separately, which means conducting a 5-classification experiment. Then, 30 (or additional 10) different images are selected from each category as the training set (or test set). In the following experiments, to accommodate the current resource configuration of the simulator and server, the dimensions of each image in the grayscale image dataset MedMNIST v2 and the color image dataset CIFAR-10 were compressed to 16 using the principal component analysis algorithm.

\subsubsection{Experimental Settings}
With a clear understanding of the dataset, the key parameters of our experiment, containing the optimizer name, learning rate, epochs, batch size and the number of qumode (or qubit), are listed in Tab. \ref{tab2}. Furthermore, since photon-type Transformers are currently rare, two state-of-the-art quantum Transformers, HQViT \cite{3.0} and Quixer \cite{3.1}, are selected as comparison objects to highlight the advanced nature of PGKET.

\begin{table}[h]
\setlength{\abovedisplayskip}{3pt}
\setlength{\belowdisplayskip}{3pt}
\def\tablename{Tab.}
\centering
\caption{Experimental Settings}
\label{tab2}
\resizebox{\columnwidth}{!}{%
\begin{tabular}{lccc}
\toprule
{Metric} & {PGKET} & {HQViT \cite{3.0}} & {Quixer \cite{3.1}} \\
\midrule
Epochs & \multicolumn{3}{c}{200} \\
Batch Size & \multicolumn{3}{c}{32 for MedMNIST v2 / 64 for CIFAR-10} \\
Learning Rate & \multicolumn{3}{c}{0.009} \\
Optimizer & \multicolumn{3}{c}{Adam} \\
\midrule
{Qmodes / Qubits} & 16 qumodes & 8 qubits & 8 qubits \\
Shots & 16 & - & - \\
\bottomrule
\end{tabular}
}
\end{table}

\subsection{Experiment 1: Noise-free 5-classification experiment}
\begin{figure*}[h!]
 \centering
 \includegraphics[width=1\textwidth]{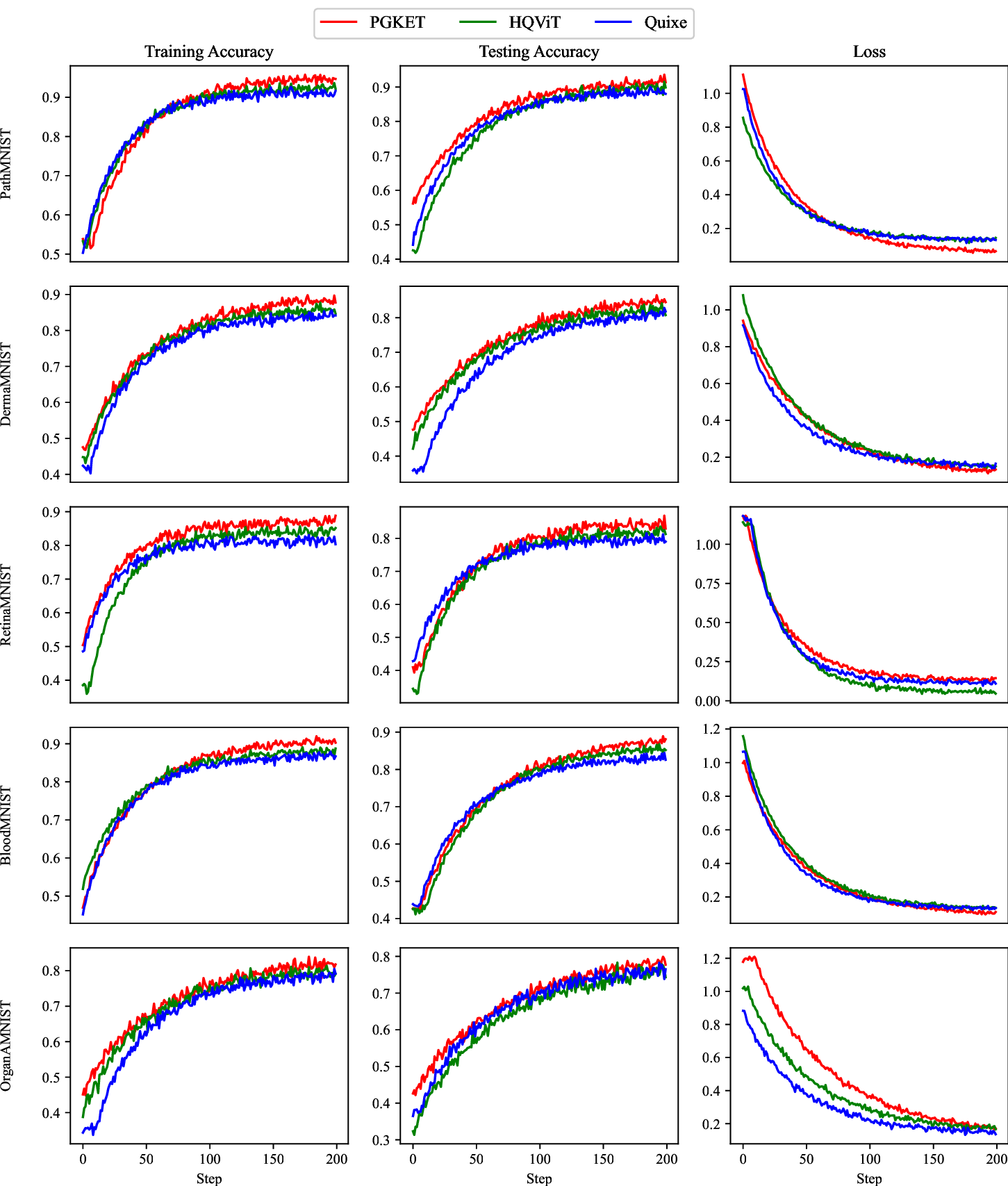}
 \caption{The noise-free multi-classification experiments for MedMNIST v2}
 \label{fig3}
\end{figure*}

\begin{figure*}
    \centering
    \includegraphics[width=1\linewidth]{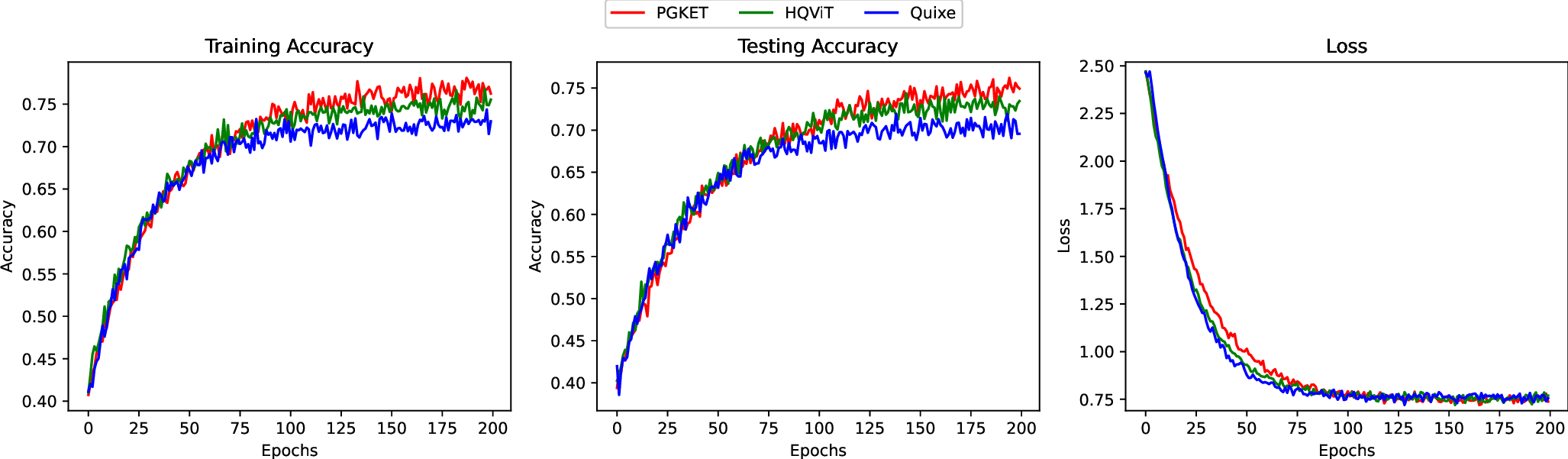}
    \caption{The noise-free multi-classification experiments for CIFAR-10}
    \label{fig4}
\end{figure*}

The noise-free multi-classification experiments for MedMNIST v2 and CIFAR-10 are shown in Figs. \ref{fig3} and \ref{fig4}, respectively. According to Figs. \ref{fig3} and \ref{fig4}, Tabs. \ref{tab3} and \ref{tab4} are obtained. 

\begin{table*}[h]
\centering
\setlength{\abovedisplayskip}{3pt}
\setlength{\belowdisplayskip}{3pt}
\def\tablename{Tab.}
\caption{Multi-classification for MedMNIST v2 summarry}
\label{tab3}
\begin{tabular}{@{}lccccc@{}}
\toprule
Dataset & PathMNIST & DermaMNIST & RetinaMNIST & BloodMNIST & OrganAMNIST \\ \midrule
PGKET Final Acc & 0.9154 & 0.8498 & 0.8448 & 0.872 & 0.7748 \\
HQViT Final Acc & 0.8955 & 0.8236 & 0.8177 & 0.85 & 0.7663 \\
Quixe Final Acc & 0.878 & 0.8107 & 0.7897 & 0.8366 & 0.7585 \\
$\Delta$Acc (PGKET-HQViT) & 0.0199 & 0.0262 & 0.0271 & 0.022 & 0.0085 \\
$\Delta$Acc (PGKET-Quixe)& 0.0374 & 0.0391 & 0.0551 & 0.0354 & 0.0163 \\
PGKET Loss & 0.1251 & 0.0916 & 0.0569 & 0.106 & 0.1787 \\
HQViT Loss & 0.1056 & 0.0713 & 0.0587 & 0.1176 & 0.1032 \\
Quixe Loss & 0.085 & 0.0836 & 0.1492 & 0.0916 & 0.1456 \\
$\Delta$Loss (PGKET-HQViT) & 0.0195 & 0.0203 & -0.0018 & -0.0116 & 0.0755 \\
$\Delta$Loss (PGKET-Quixe) & 0.0401 & 0.0080 & -0.0923 & 0.0144 & 0.0331\\
PGKET Conv Epoch & 109 & 130 & 94 & 115 & 147 \\
HQViT Conv Epoch & 101 & 114 & 104 & 118 & 134 \\
Quixe Conv Epoch & 87 & 110 & 88 & 106 & 135 \\
$\Delta$ Epoch (PGKET-HQViT) & 8 & 16 & -10 & -3 & 13 \\
$\Delta$ Epoch (PGKET-Quixe) & 22 & 20 & 6& 9 & 12 \\ \bottomrule
\end{tabular}
\end{table*}
\begin{table*}[h]
\centering
\setlength{\abovedisplayskip}{3pt}
\setlength{\belowdisplayskip}{3pt}
\def\tablename{Tab.}
\caption{Multi-classification for CIFAR-10 summary}
\label{tab4}
\begin{tabular}{@{}lccccc@{}}
\toprule
Metric & PGKET & HQViT & Quixe & $\Delta$Acc (PGKET-HQViT) & $\Delta$Acc (PGKET-Quixe) \\ \midrule
Final Test Acc & 0.7494 & 0.7345 & 0.6957 & 0.0149 & 0.0537 \\
Final Loss     & 0.7379 & 0.7689 & 0.7555 & -0.0310 & -0.0176 \\
Convergence Epoch & 101 & 97 & 83 & 4 & 18 \\ \bottomrule
\end{tabular}
\end{table*}

For MedMNIST v2, the following conclusion can be drawn: (1) Accuracy Advantage: Across all five datasets, PGKET demonstrates a consistent performance improvement, achieving an average increase of approximately 2.09\% over HQViT and 3.64\% over Quixe. The most substantial gain occurs on RetinaMNIST, where PGKET outperforms HQViT by 2.71\% and Quixe by 5.51\%, suggesting superior efficacy on this complex multi-class classification task. The smallest improvement is observed on OrganAMNIST, with PGKET surpassing HQViT by 0.85\% and Quixe by 1.63\%, indicating more modest gains on simpler, smaller-scale tasks.
(2) Loss Stability: The final loss for PGKET averages 0.111 across the five datasets, comparable to Quixe (0.111) but slightly higher than HQViT (0.091). In specific datasets such as PathMNIST, DermaMNIST, and BloodMNIST, PGKET's loss is marginally higher than HQViT by approximately 0.017 on average, yet it remains lower than that of Quixe. In contrast, on more challenging datasets like RetinaMNIST and OrganAMNIST, PGKET significantly outperforms Quixe, with loss reductions of -0.0923 and -0.0331, respectively, indicating better convergence to lower loss in these instances.
(3) Convergence Speed: On average, PGKET converges to 95\% of the peak accuracy in 119 epochs, while HQViT converges in 114 epochs, and Quixe in 105 epochs. PGKET demonstrates a slightly slower convergence on PathMNIST (+8 epochs) and DermaMNIST (+16 epochs), but outperforms or matches the other models on RetinaMNIST (-10 epochs), BloodMNIST (-3 epochs), and OrganAMNIST (+13 epochs). This suggests that PGKET offers a balanced trade-off between convergence speed and performance, achieving comparable or superior results without compromising efficiency.
Overall Conclusion: PGKET consistently achieves the highest final accuracy across all test sets, with improvements ranging from 1\% to 4\%, fulfilling its design objectives. In terms of loss, PGKET exhibits superior performance on high-difficulty datasets (RetinaMNIST, OrganAMNIST), reflecting its better convergence quality. Regarding convergence speed, while PGKET slightly lags behind HQViT on certain datasets, its overall convergence epoch is competitive with Quixe, and when considering its accuracy advantage, PGKET demonstrates superior overall efficiency.

Another conclusion for CIFAR-10 is: (1) Accuracy Advantage: PGKET achieves a final test accuracy of 0.7494 on the CIFAR-10 dataset, representing improvements of 1.49\% and 5.37\% over the 0.7345 of HQViT and the 0.6957 of Quixe, respectively, demonstrating a significant advantage in classification performance. (2) Loss Stability: The final loss value of PGKET is 0.7379, lower than that of HQViT (0.7689) and Quixe (0.7555), representing reductions of 0.0310 and 0.0176, respectively, indicating superior convergence quality and prediction reliability. (3) Convergence Speed: PGKET requires 101 epochs to reach a peak accuracy of 95\%, slightly longer than HQViT (97) and Quixe (83), but its higher accuracy and lower loss reflect a better balance between performance and efficiency. (4) Overall conclusion: PGKET outperforms HQViT and Quixe in terms of accuracy and loss, despite slightly increased convergence time. Its improved overall performance validates its effectiveness as an advanced model for image classification.

By comparing Tabs. \ref{tab3} and \ref{tab4}, the PGKET model demonstrates exceptional accuracy and strong generalization capabilities across a wide range of medical and natural image datasets, combining robustness with training efficiency, thereby showcasing its potential as a general-purpose visual classification model. Specifically, (1)The efficient performance of PGKET on diverse medical image datasets: In the MedMNIST v2 dataset, PGKET achieved high classification accuracy (ranging from 0.7748 to 0.9154) and low loss values, with a relatively moderate training cycle, demonstrating its strong generalization ability and stability in the field of medical imaging. This indicates that PGKET is suitable for handling medical imaging tasks with relatively complex structures and diverse categories.
(2)The robustness and competitiveness of PGKET in natural image datasets: In the CIFAR-10 dataset, a typical natural image dataset, PGKET also achieved a high accuracy rate (0.7494), outperforming HQViT and Quixe, with lower loss and a reasonable training convergence cycle. This shows that PGKET is not only adaptable to medical image tasks but also has good cross-domain adaptability.
(3) Balanced cross-domain performance and generalization potential: PGKET demonstrates stable performance on both medical image and natural image datasets, indicating that its design has strong generalization capabilities, can adapt to different types and complexities of data features, and has high application value.
(4) Balanced training efficiency: Althoughthe convergence epoch (101) of PGKET in CIFAR-10 is slightly higher than that of HQViT (97), it remains within a reasonable range and is slightly longer than that of Quixe (83), resulting in balanced overall training efficiency. Considering the training epoch on MedMNIST, PGKET achieves a good trade-off between accuracy and training time.

\subsection{Experiment 2: Noise Robustness Test}

\begin{figure*}[h!]
    \centering
    \includegraphics[width=0.95\linewidth]{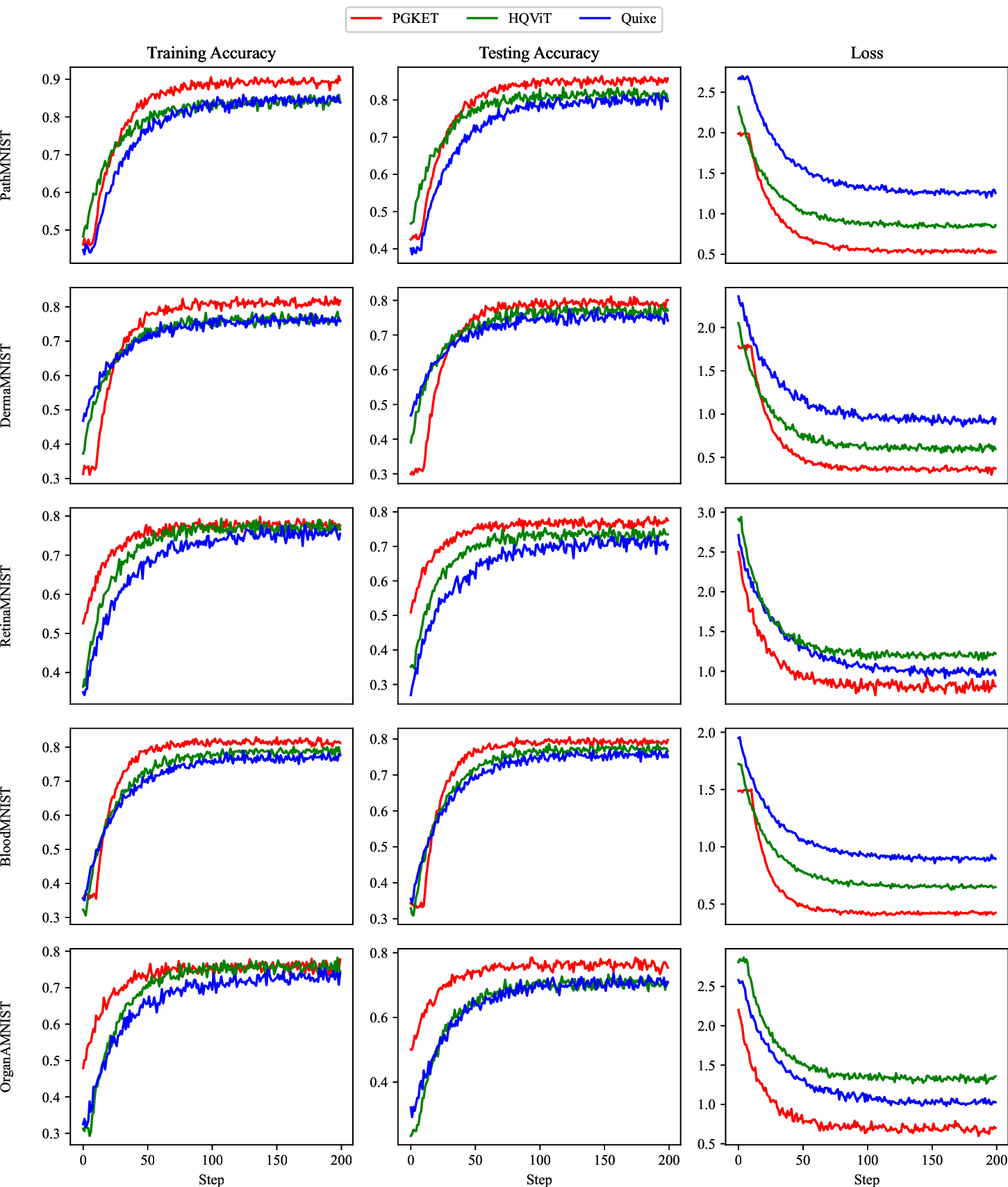}
    \caption{The noisy multi-classification experiments for MedMNIST v2}
 \label{fig5}
\end{figure*}

Here, 0.4 Gaussian noise is applied to each of the 5 datasets in MedMNIST v2. Then, training is performed under the same configuration as in Experiment 1. Finally, Fig. \ref{fig5} and Tab. \ref{tab5} are obtained.
\begin{table*}[h!]
\centering
\setlength{\abovedisplayskip}{3pt}
\setlength{\belowdisplayskip}{3pt}
\def\tablename{Tab.}
\caption{Noisy 5-classification Performance Summary for MedMNIST v2}
\label{tab5}
\begin{tabular}{@{}lccccc@{}}
\toprule
Dataset & PathMNIST & DermaMNIST & RetinaMNIST & BloodMNIST & OrganAMNIST \\ \midrule
PGKET Final Acc & 0.8532 & 0.7590 & 0.7564 & 0.8327 & 0.7650 \\
HQViT Final Acc & 0.8166 & 0.7478 & 0.7122 & 0.8087 & 0.7076 \\
Quixe Final Acc & 0.7996 & 0.7395 & 0.6967 & 0.8017 & 0.7077 \\
$\Delta$Acc (PGKET-HQViT) & 0.0366 & 0.0112 & 0.0442 & 0.0240 & 0.0574 \\
$\Delta$Acc (PGKET-Quixe) & 0.0536 & 0.0195 & 0.0597 & 0.0310 & 0.0573 \\
PGKET Loss & 0.5376 & 0.5369 & 0.5338 & 0.4877 & 0.6809 \\
HQViT Loss & 0.5350 & 0.6457 & 0.5836 & 0.5330 & 0.6598 \\
Quixe Loss & 0.5578 & 0.6453 & 0.6127 & 0.4474 & 0.6479 \\
$\Delta$Loss (PGKET-HQViT) & 0.0026 & -0.1088 & -0.0498 & -0.0453 & 0.0211 \\
$\Delta$Loss (PGKET-Quixe) & -0.0202 & -0.1084 & -0.0789 & 0.0403 & 0.0330 \\
PGKET Conv Epoch & 100 & 80 & 90 & 70 & 85 \\
HQViT Conv Epoch & 120 & 110 & 100 & 115 & 105 \\
Quixe Conv Epoch & 140 & 130 & 150 & 135 & 145 \\
$\Delta$Epoch (PGKET-HQViT) & -20 & -30 & -10 & -45 & -20 \\
$\Delta$Epoch (PGKET-Quixe) & -40 & -50 & -60 & -65 & -60 \\ \bottomrule
\end{tabular}
\end{table*}
According Fig. \ref{fig5} and Tab. \ref{tab5}, PGKET demonstrated better accuracy, lower loss, shorter training cycles, and better training stability compared to the other two models in noisy environments, indicating its stronger robustness and efficiency when dealing with complex data and noise. Specifically, (1) Higher classification accuracy: PGKET achieves higher final accuracy across all datasets, particularly on the PathMNIST (0.8532) and BloodMNIST (0.8327) datasets, where it outperforms HQViT (0.8166, 0.8087) and Quixe (0.7996, 0.8017) by a significant margin. The improved accuracy of PGKET compared to other models indicates its ability to maintain performance in noisy environments. (2) Lower loss values:The loss values of PGKET are generally lower than those of HQViT and Quixe, especially on datasets such as PathMNIST (0.5376) and RetinaMNIST (0.5338), where the loss values of PGKET are significantly better than those of the other two models (HQViT: 0.5350, 0.5836; Quixe: 0.5578, 0.6127). This reflectsthe effectiveness of PGKET in handling noise, enabling it to better fit the data. (3) Training cycle advantage: PGKET demonstrates shorter training cycles (epochs) compared to HQViT and Quixe across multiple datasets, particularly on the BloodMNIST dataset, where PGKET requires 70 training cycles, significantly fewer than HQViT (115 epochs) and Quixe (135 epochs). This indicates that PGKET has a notable advantage in computational efficiency.
(4) Relatively small performance degradation: Compared to HQViT and Quixe, PGKET exhibits a relatively smaller decrease in accuracy, particularly on the PathMNIST (0.0366) and BloodMNIST (0.0574) datasets, where PGKET demonstrates stronger robustness compared to HQViT and Quixe. Even in noisy environments,the performance loss of PGKET is far smaller than that of other models.
(5) Training stability: PGKET demonstrates better stability during training compared to other models. Especially in terms of loss function decline, PGKET exhibits a smoother change in loss compared to HQViT and Quixe, particularly on the RetinaMNIST (-0.0498) and OrganAMNIST (0.0211) datasets, where PGKET demonstrates relatively good training stability.

\section{Conclusion}\label{sec5}
This paper proposes a new Transformer architecture, PGKET, whose core innovation is to introduce the photon Gaussian kernel self-attention mechanism (PGKSAM), combining PC and kernel methods to improve the efficiency of long sequence processing and model performance. PGKSAM achieves efficient and parallel attention score calculation through photon interference and superposition, replacing the traditional softmax mechanism and significantly reducing the computational complexity.
A large number of experiments have shown that PGKET outperforms existing quantum Transformer models (HQViT and Quixe) in multi-classification tasks on the MedMNIST v2 and CIFAR-10 datasets, showing advantages in accuracy, loss stability and convergence speed. Especially in noisy environments, PGKET still maintains high robustness and training efficiency, verifying its practicality and prospects in complex data processing.

\end{document}